\documentclass[aps,prb,twocolumn]{revtex4}
\usepackage{graphicx}

\begin{document}

\title{Study on Unconventional Superconductors via Angle-resolved Specific Heat}

\author{Tuson Park}
\address{Los Alamos National Laboratory, Los Alamos, New Mexico 87545}

\author{M. B. Salamon}
\address{Department of Physics and Materials Research Laboratory, University of
Illinois at Urbana--Champaign, Urbana, Illinois 61801}

\date{\today}

\begin{abstract}
The gap function in unconventional superconductors may vanish at
points or lines in momentum space, permitting electronic
excitations, termed nodal quasiparticles, to exist at temperatures
well below the superconducting transition. In the vortex phase,
the presence of nodal quasiparticles should be directly observable
through the variation of the heat capacity with the angle between
a magnetic field and the location of the zeroes of the gap. The
heat capacity of candidate non-magnetic unconventional superconductors Lu(Y)Ni$_{2}$B$_{2}$C were found to exhibit
fourfold oscillations with field angle, the first such
observation. The observed angular variations are in quantitative
agreement with theory, confirming that quasiparticles are created
via Doppler shifts at nodes along $<100>$. Anomalous disorder effects have been also observed in the field-angle dependent heat capacity $C_{p}(\alpha)$. In a slightly disordered sample, anomalous secondary minima along $<110>$ appeared for $\mu _{0}H > 1$~T, leading to an eightfold pattern. The coexistence of an anisotropic superconducting gap and nonlocal effects is shown to drive the anomalous behavior. These results demonstrate that field-angle-dependent heat capacity can be a powerful tool in probing the momentum-space gap structure in
unconventional superconductors such as high T$_{c}$ cuprates, heavy fermions,
borocarbides, etc.
\end{abstract}

\maketitle

\section{Introduction}
Most superconductors behave conventionally; electronic
excitations are suppressed by the BCS gap causing the electronic
heat capacity, for example, to be exponentially small at
temperatures well below the superconducting transition. Recently,
a number of superconductors, cuprates \cite{annett96} and
heavy-fermion metals \cite{sigrist91} among them, have been found
to be unconventional in that they exhibit gap-zero (or nodal)
points or lines in momentum space. Electronic excitations - nodal
quasiparticles (nqp) - are then observed at low temperatures,
giving rise to power-law rather than exponential, behavior. Unlike
gapless superconductivity, which can occur in conventional
superconductors, the Fermi momenta of these quasiparticles are
restricted to nodal regions of the Fermi surface, giving a strong
directional dependence to various physical properties.

In 1992, Yip and Sauls proposed a nonlinear Meissner effect in the
penetration depth of cuprate high-$T_{c}$ superconductors.
\cite{yip92} Since the effect uses the field-direction dependence
of the supercurrent in locating the positions of the nodal lines
(or points) of an unconventional gap in momentum space, it is a
stronger test of the symmetry of the superconducting state than
power-law behavior for the penetration depth. The search for the
fundamental manifestation of d-wave symmetry has been very
controversial from an experimental point of view.
\cite{maeda95,maeda96,bhattacharya99,carrington99,bidinosti99} Li
\textit{et al.} questioned the observability of the nonlinear Meissner
effect.\cite{li98} They introduced nonlocal electrodynamics for a
d-wave superconductor and found that the local nonlinear Meissner
effect is not observable for fields below a crossover scale
$H^{\ast }$ describing the competition between nonlinear and
nonlocal effects in the Meissner state. For most orientations of
the screening current, the crossover is of the same order as or
greater than the lower critical field $H_{c1}$.

In conventional type II superconductors, the vortex cores play a
crucial role in transport properties because a relatively high
density of bound states is created there by the suppression of the
local order parameter and extended quasiparticle states are
completely gapped. In a new symmetry class such as cuprate
superconductors, however, Hirschfeld \textit{et al.} argued that extended
quasiparticle states are dominant over vortex bound states because
of the presence of the nodes.\cite{hirschfeld98} The anisotropy
in the gap $\Delta _{k}$ in momentum space fixed to the crystal
axes will induce an angular anisotropy in the current response due
to the coupling between quasiparticles and supercurrent, leading
to the fourfold angular variation in field-angle thermal
conductivity of unconventional superconductors with d-wave order
parameter. In 1995, Yu and Salamon \textit{et al.} demonstrated for the
first time that the heat transport of YBCO oscillates with
in-plane field angle.\cite{salamon95,yu95} Subsequent experiments
have confirmed the feature and thermal conductivity has been
established as a tool that can probe the momentum-specific gap
nature in unconventional superconductors.\cite{aubin97,izawa01,izawa02} Such measurements, however, are
complicated by the competition between Andreev-scattering
and Doppler-shift effects.

A much more direct, and far more difficult, experiment is to
detect the low-energy density of states of the nodal quasiparticles through
its modulation by an in-plane field. Heat capacity provides
such a measure. Working in the 2D limit, Vekhter \textit{et al.} have shown
that the density of states of a d-wave superconductor exhibits
four-fold oscillation with field angle with respect to crystal
axes.\cite{vekhter99} From the experimental stand point, the search
for the oscillation has been puzzling.\cite{moler94,wang01} The
expected 30~\% of oscillation amplitude is within the reach of
experimental error, but the effect has proven elusive. Whelan
and Carbotte argued that the Zeeman energy splitting $\mu H$
competes against the anisotropy.\cite{whelan00} At a critical
field $H_{c}$, Zeeman and Doppler energy scales become comparable
and the DOS variation with field angle vanishes. Recently, Won and
Maki used a semi-classical approximation and extended the 2D
Vekhter model to a 3D superconductor with modulated cylindrical
Fermi surface by including quasiparticles with an out-of-plane momentum
component.\cite{won01} The 3D effect causes a strong suppression
of the oscillation amplitude to around 6~\%, making the
observation much more difficult.

In this paper, we review the direct observation of a
phenomenon that should be common to many so-called unconventional
superconductors, those whose Cooper pairs have symmetries more
complex than originally envisaged by Bardeen, Cooper and Shrieffer
in their classic theory of superconductivity. If, in particular,
the pairs have the symmetry similar to d-orbitals in atoms, there
will be certain directions in momentum space in which electronic
excitations, termed quasiparticles, can arise without an energy
gap.

In section~2, the experimental method, sample
preparation and characterization will be briefly explained. In section~3,
the field-angle heat capacity of LuNi$_{2}$B$_{2}$C is shown to be similar to results reported previously for YNi$_{2}$B$_{2}$C.\cite{tuson03} The magnetic field dependence will be analyzed by a Won-Maki extension of the Vekhter \textit{et al.} calculation. The heat capacity dependence on magnetic field
angle will be discussed in terms of 3D nodal quasiparticle theory
and used as a piece of confirming evidence that LuNi$_{2}$B$_{2}$C and YNi$_{2}$B$_{2}$C belong to a class of superconductors with nodes in gap function. An eightfold pattern appeared in a slightly disordered Lu1221 supports the coexistence of nonlocal effects and anisotropic gap effects. Then, concluding remarks will follow in section 4.

\section{Experiments}
The electronic specific heat is a fundamental physical quantity that
measures the electronic density of states (DOS) directly and is a bulk property.
As noted in the Introduction, it was recently suggested that the specific heat can be used to study the angle-resolved gap structure of unconventional superconductors by studying the DOS variation with magnetic field angle.\cite{vekhter99} Considerable efforts have been expended to
detect the field-angle variation arising from the anisotropic gap
structure in unconventional superconductors, especially in YBCO.\cite{moler94,wang01} Those efforts, however, have suffered from
intrinsic experimental limitations. First, a very limited set of
field directions were investigated. Typically, the crystal-axes
and diagonal directions were chosen because they give the largest
contrast in DOS for pure $d_{x^{2}-y^{2}}$ or $d_{xy}$ pairing symmetry. If the pairing symmetry is the mixture of two order parameters, such as
$d+s-$wave, then the two directions do not represent gap-maximum
or gap-zero (nodes) directions because the nodes move away from
the diagonal directions. Second, temperature-variable heat
capacity intrinsically contains large background contributions
such as lattice vibrations or thermally excited quasiparticles.
The large backgrounds cause an uncertainty in extracting the
field-induced part, and can easily obscure the small angular
variation.

\begin{figure}[tbp]
\centering  \includegraphics[width=8.5cm,clip]{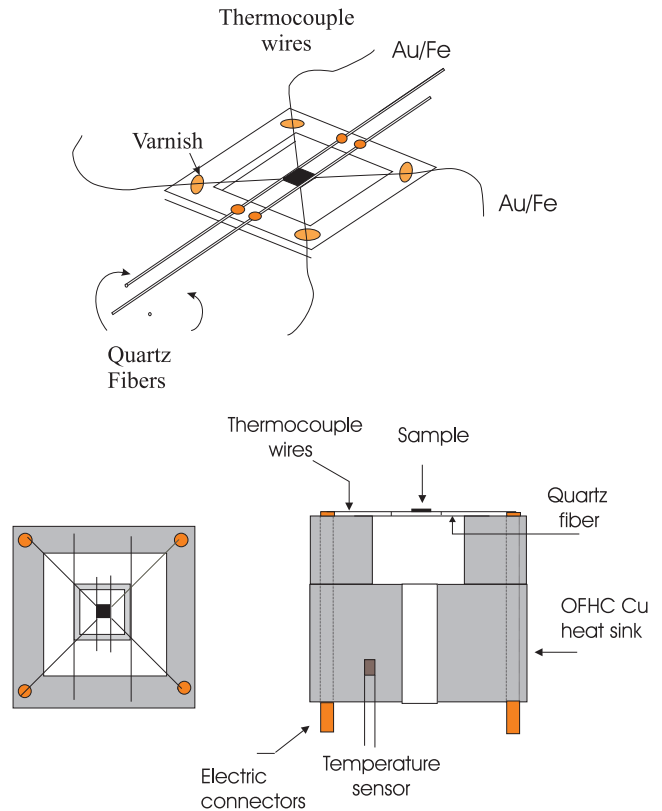}
\caption{Sample assembly in ac calorimetry}
\end{figure}
With those problems in mind, we built a new specific heat probe
designed exclusively to study the gap structure of superconductors
in momentum space. Instead of varying temperature as was done in
many previous works, we fixed temperature and magnetic field
intensity and made the field angle as an only variable. This
experimental scheme enabled us to study the small angular
oscillation, which is typically less than 1 $\%$ of the measured
heat capacity. Some of the key ingredients to this scheme are as
following. First, temperature and magnetic field should be very
stable. Otherwise, even a small fluctuation could obscure the small
change with angle. A Physical Properties Measurement System (PPMS)
by Quantum Design was used as a cryogenics platform to satisfy those
conditions. Second, a stepping motor with a gear ratio 1:141 was
used to rotate the sample and was controlled by a driver from
PPMS. Instead of measuring limited directions, we were able to
study the angle dependence of heat capacity in increments of
3$^{\circ }$ or smaller. Third, the heat capacity measurement
should be sensitive enough to pick up the minute change with field
angle. In addition, the measurement must be able to handle small
single crystals because their weight is usually less than
1~mg. AC calorimetry was chosen because it has a sensitivity of
up to 0.01 $\%$ of the measured heat capacity as well as a
capability to measure even microgram ($\mu g)$ samples.\cite{kraftmakher02}

\begin{figure}[tbp]
\centering  \includegraphics[width=8.5cm,clip]{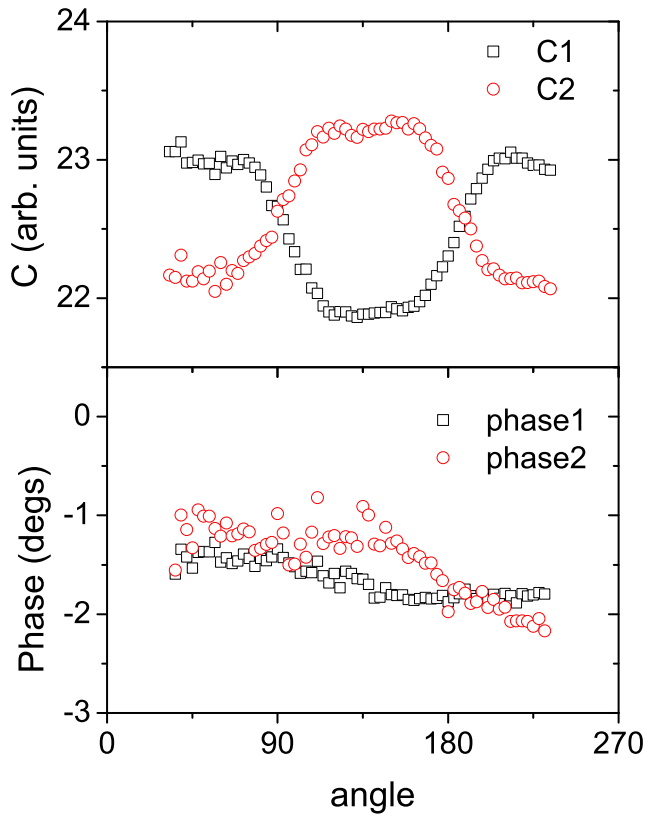}
\caption{Upper panel shows the field-angle heat capacity of
optimally doped YBCO at 1.7~K and 7~T. The phase shift with field
angle is shown in the lower panel. In both  panels, the squares
describe the data measured by the first pair of thermocouple wires
and the circles by the second pair.}
\end{figure}
The $ac$ method used here is an adaptation of the Sullivan and Seidel
technique.\cite{sullivan68} Fig.~1 shows the sample assembly. The front face of the sample was coated with colloidal graphite suspension (DAG) thinned with
isopropyl alcohol to prevent a possible change of the optical
absorptions of the sample. The sample was weakly
coupled to the heat bath through helium gas and suspending
thermocouple wires. At high temperatures, the thermal coupling can
be adjusted mainly by controlling the amount of the helium gas
surrounding the sample, while the thermal leak through thermocouple
wires should be taken into account at cryogenic temperatures. As
a heating source, infrared laser light ($\lambda =789.2$ nm) was
used. The light was modulated electronically to make a square-wave
pulse and was guided into the PPMS sample space through an optical fiber.
The oscillating heat input incurred a steady temperature offset (or $dc$ offset) against the heat bath with an oscillating temperature superposed. The measured
oscillation temperature $T_{ac}$ was converted to heat capacity by the relationship $C \propto 1/T_{ac}$ in a proper frequency range.

In our field-angle heat capacity measurements, two thermocouple
pairs were used to measure the temperature oscillations $T_{ac}$
simultaneously. The two sets of Au/Fe and chromel thermocouple wires were configured so that they were 90$^{\circ }$ apart, which enabled us to rule out the field-angle contribution from the thermocouple wires due to field-dependent thermopower of Au/Fe. If the angular oscillations come from our experimental setup, the two signals will be out of phase while
they will be in phase with each other if they come from sample. As
an example, we studied an optimally doped YBCO (Fig.~2). The squares in the top panel describe the angular heat capacity measured by the first pair of the
thermocouple wires and the circles by the second pair. The two
sets of data were measured simultaneously at 1.7~K in 7~T with
magnetic field rotating within the CuO$_{2}$ plane of YBCO. Both
of them show clear two-fold oscillations with field angle measured
against a-axis. Note that the oscillations are out of phase with
each other, indicating that they come from the thermocouple wires. The
bottom panel of Fig.~2 shows the phase shift with field angle.
There is no correlation with the heat capacity shown in the upper
panel, confirming that the two-fold oscillation of
the heat capacity is not related to intrinsic sample properties.\cite{thesis}

\begin{figure}[tbp]
\centering  \includegraphics[width=8.5cm,clip]{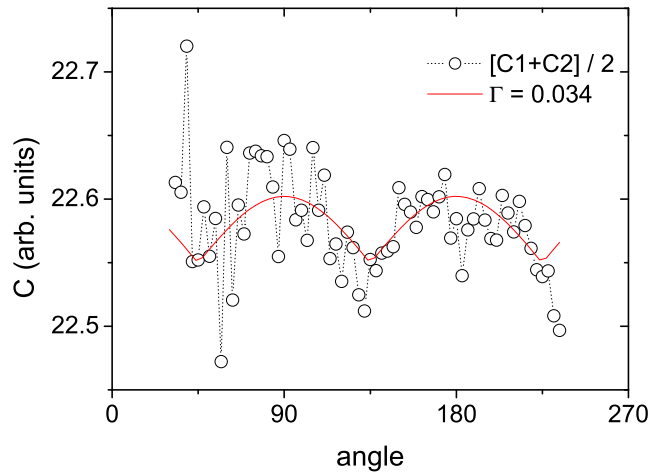}
\caption{The averaged field-angle heat capacity of YBCO at 1.7~K
and 7~T. The solid line shows a theoretical calculation of fourfold
oscillation with the oscillation amplitude of 3.4~$\%$ (see text).}
\end{figure}
In Fig.~3, we show the averaged heat capacity of YBCO
($2C_{ave}=C_{1}+C_{2}$). As expected, the two-fold components are
almost suppressed. Now that the artifact from the experimental
setup is removed, we can study the intrinsic sample property. The
solid line describes a fit by a cusped fourfold variation with
oscillation amplitude of 3.4 $\%$, i.e., $C(\alpha )=C_{0}+c_{4}(1+\Gamma |\sin 2\alpha |)$. Here, $\alpha$ is the magnetic field angle with respect to the $a$-axis and $\Gamma$ is the oscillation amplitude. The periodicity is $\pi /4$ and the minima are located at $\langle 110\rangle $ directions, which
are consistent with $d_{x^{2}-y^{2}}-$wave order parameter in YBCO
where the nodes are located along $\langle 110\rangle $. The
field-angle oscillation, however, is obscured by noise which is
comparable to the angular change.

\section{Observation of nodal quasiparticles and nonlocal effects in the nonmagnetic superconductors Lu(Y)Ni$_{2}$B$_{2}$C}

The discovery of RNi$_{2}$B$_{2}$C (R = Y, Lu, Tm, Er, Ho, and Dy)
captured attention of the superconductivity community
because of its relatively high transition temperature $(T_{c})$
and the coexistence of magnetism and superconductivity.\cite{cava94,mazumdar93,canfield98} The highest $T_{c}$'s in the borocarbide family are 16.5 K and 15.5 K in the nonmagnetic
members with Lu and Y elements respectively. Amid the
controversy over the pairing symmetry of the borocarbides, there
is growing evidence that the gap function is highly anisotropic
and possibly has nodes where the gap becomes zero in momentum
space. In specific heat measurements, a power law behavior was
observed in the temperature dependence and the electronic
coefficient $\gamma (H)$, to the extent it can be extracted,
follows a square-root field dependence.\cite{nohara97,volovik93}
Thermal conductivity measurements showed low energy excitations at
as low as 70 mK and an anomalous magnetic field dependence
similar to the unconventional superconductor UPt$_{3}$ but
different from the exponential dependence characteristic of an isotropic
s-wave gap.\cite{boaknin01} Compelling evidence for the
presence of nodes along $\langle 100\rangle $ directions has been
reported from both z-axis field-angle thermal conductivity
\cite{izawa02} and field-angle heat capacity measurements of
YNi$_{2}$B$_{2}$C.\cite{tuson03}

Recently, a magnetic field-driven flux line lattice (FLL) transition has been observed both in the magnetic member (Er) and in the nonmagnetic members (Y and Lu) of the tetragonal borocarbides.\cite{yaron96,eskildsen98,sakata00,eskildsen2001}
The transition from square to hexagonal vortex lattice occurs due to the
competition between sources of anisotropy and vortex-vortex interactions. The
repulsive nature of the vortex interaction favors the hexagonal Abrikosov
lattice, whose vortex spacing is larger than that of a square lattice. The
competing anisotropy, which favors a square lattice, can be due to lattice
effects (fourfold Fermi surface anisotropy) \cite{kogan97} or unconventional
superconducting order parameter.\cite{gilardi02} In this section, we present evidence that both the gap anisotropy and nonlocal effects coexist and the coexistence is the origin of the anomalous crossover from a fourfold pattern to an eightfold pattern in $C(\alpha)$ with increasing field in LuNi$_{2}$B$_{2}$C.

\subsection{Disorder effects}

The best samples of LuNi$_{2}$B$_{2}$C exhibit behavior identical to that reported earlier for YNi$_{2}$B$_{2}$C,\cite{tuson03} but disordered samples are significantly different.
Three single crystals of LuNi$_{2}$B$_{2}$C from different batches were studied. Sample N was annealed at $T=1000$~C$^{\circ}$ for 100 h under high vacuum, while sample A and C were not annealed. The annealed crystal, sample~N, showed the highest $T_{c}$ (16.1~K) and residual resistivity ratio (RRR) of 27, while the unannealed sample~A recorded the lowest $T_{c}$ of 15.5~K and RRR of 16 (see Fig.~4), suggesting a judicious postgrowth annealing may improve the sample quality. The resistivity at $T_{c}$ is 2.34 and 1.44~$\mu \Omega \cdot$cm for samples~A and~N, corresponding to mean-free paths of 144.5 and 234~$\mathring{A}$ respectively. Assuming that 16.1~K is the $T_{c}$ for a pure sample, sample~A is equivalent to 0.8~\% of Co doping on the Ni site; i.e., Lu(Ni$_{1-x}$Co$_{x}$)$_{2}$B$_{2}$C with  $x=0.008$.\cite{cheon98} Sample~C with $T_{c}=15.8$~K is less disordered than sample~A.
\begin{figure}[tbp]
\centering  \includegraphics[width=8.5cm,clip]{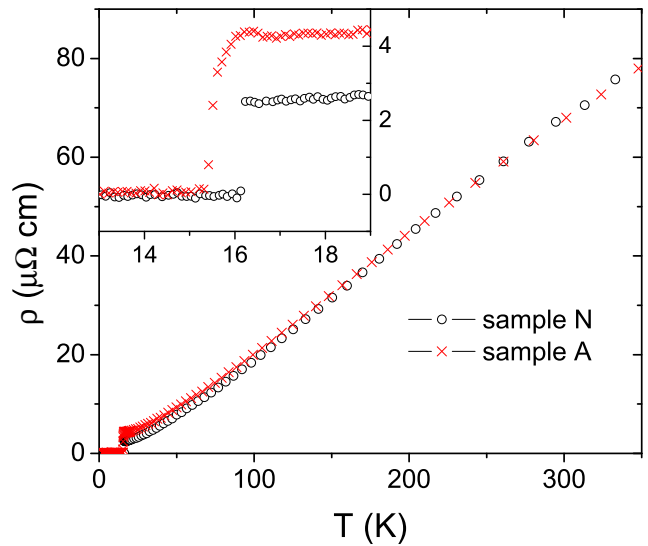}
\caption{Temperature dependence of the in-plane resistivity of sample~A (crosses) and sample~N (circles).}
\end{figure}

For an isotropic s-wave superconductor, the contribution to the density of states (DOS) comes from the localized states in the vortex core. Since the density of vortices is proportional to the magnetic field, the field-induced DOS is linearly dependent on the field and $\beta =1$. For superconductors with line nodes, delocalized states leaking through the nodes dominate the DOS. When we
integrate the contribution in a vortex unit cell and multiply by the
density of vortices, the electronic DOS is proportional to
$\sqrt{H}$ in an intermediate region, i.e. $H_{c1}\ll H \ll
H_{c2}$, where $H_{c2}$ is upper critical field. This so called Volovik effect gives $\beta =0.5$ for line node and $\beta =1$ for point node.\cite{volovik93} Recently, there have been several reports that s-wave superconductors such as
CeRu$_{2}$ \cite{hedo98} and NbSe$_{2}$ \cite{nohara97} behave
like d-wave superconductors, showing sub-linear magnetic field
dependence in specific heat. Ichioka \textit{et al.}\cite{ichioka99} extended the Volovik calculation to the entire region, i.e. $H_{c1} <H<H_{c2}$. While the
Volovik effect considered the delocalized states only, they included the localized states bound in a vortex core and the quasiparticle transfers between vortices. In their numerical calculation, the DOS dependence in the d-wave pairing states gives an exponent $\beta =0.41$, rather than Volovik's $\beta =0.5$. As for those with anisotropic s-wave pairing symmetry, the exponent becomes 0.67, which is clearly distinguishable from 0.5 or 0.41.

The disorder effects are also manifested in the magnetic field dependence of the electronic specific heat $C_{el}$. In pure samples (sample~C and N), the low-temperature specific heat shows a $\sqrt{H}$ dependence in the mixed state, but there is a deviation from the $\sqrt{H}$ dependence at a finite field in the disordered sample~A. Fig.~5(a) shows the heat capacity of sample~C at 2.5~K with increasing (circles) and decreasing (crosses) magnetic field along [100] direction. The dashed line which represents the least square fit of $C_{0}+b(H-H_{0})^{\beta}$ best describes the field dependence with $\beta=0.46$. Here $C_{0}$ is the zero-field heat capacity and the offset $H_{0}$ is essentially a lower critical field ($H_{c1}$) and 0.1~T was used for the least-square fit.
\begin{figure}[tbp]
\centering  \includegraphics[width=8.5cm,clip]{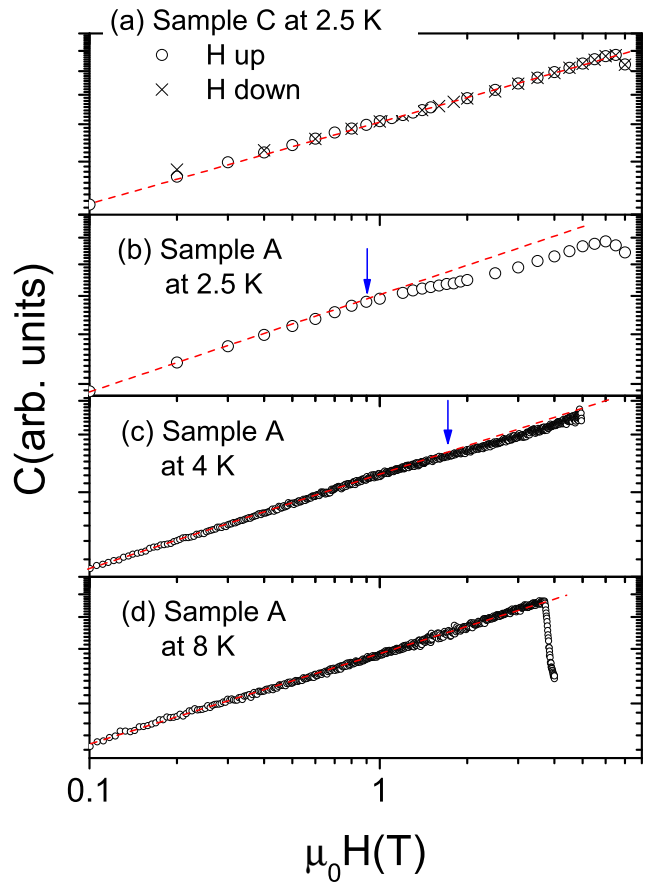}
\caption{(a) Heat capacity of sample~\textit{C} at 2.5~K for $H \parallel$[110]. (b)-(d) Heat capacity of sample~\textit{A} at 2.5, 4, and 8~K, respectively for $H \parallel$[110]. Dashed lines are $H^{1/2}$ fits and arrows indicate the deviation points from the $H^{1/2}$ dependence.$^{38}$}
\end{figure} 

Figures~5(b)-(d) show the heat capacity of the disordered sample~A at 2.5, 4, and 8~K respectively. Low-field data were best described by the square-root field dependence (dashed lines) and the arrows indicate the points where the data deviate from the fit. The deviation field shows a systematic increase with temperature, i.e, 0.8~T at 2.5~K, 1.8~T at 4~K, and no clear deviation at 8~K. Above the deviation field, the data fall below the $\sqrt{H}$ line.

\subsection{Angle-resolved Specific Heat}

In addition to the $\sqrt{H}$ dependence in unconventional superconductors, the density of states $N(H,\alpha)$ also depends on the orientation of the superfluid flow with respect to the nodes, which results in angular oscillation in $C(\alpha)$.\cite{vekhter99,miranovic03} The supercurrent flow around a vortex leads to a Doppler energy shift, $\delta E = \mathbf{v}_{s}\cdot \hbar \mathbf{k}_{F}$ , where
$\mathbf{v}_{s}$ is the velocity of the superfluid and $\hbar \mathbf{k}
_{F}$ is the Fermi momentum of nodal quasiparticles. When the
field direction is normal to the plane containing nodes, the DOS
is the average over the whole Fermi surface, leading to a square
root field dependence. When the field is in the nodal plane,
however, the Doppler shift has a field-direction dependence as
well, $\delta E \approx \frac{E_{h}}{\rho }\sin \beta \sin
(\phi -\alpha )$.\cite{vekhter99} Here $\phi $ is an azimuthal
angle of the gap node and $\beta $ is a vortex current winding
angle. The energy scale associated with the Doppler effect is
defined as
\begin{equation}
E_{h}=\frac{a\hbar v^{\ast }}{2}\left( \frac{\pi H}{\Phi
_{0}}\right)^{1/2},
\end{equation}
where the geometrical constant $a$ is order of unity, $v^{\ast
}=\sqrt{v_{a}v_{c}}$ is the average Fermi velocity, and $\Phi
_{0}$ is the flux quantum. The variable $\rho =r/R$, where $r$ is the distance from the vortex core and $2R$ is the intervortex distance. Working in the 2D limit, Vekhter \textit{et al.} calculated the DOS , $N\approx (N_{1}+N_{2})/2$, for an in-plane
magnetic field when $\omega ,E_{h}\ll \Delta _{0}$ where $\omega$ is the Matsubara frequency, and for four nodes at angles $\alpha _{n}$ from orthogonal axes in a plane :
\begin{equation}
\frac{N_{i}(\omega ,h,\alpha )}{N_{0}}=\left\{
\begin{array}{ll}
\frac{\omega }{\Delta _{0}}(1+\frac{1}{2x^{2}}) \hspace{1.7cm} (x=\omega /E_{i}\geq 1) \\
\frac{E_{i}}{\pi \Delta _{0}x}[(1+2x^{2})\arcsin x+3x\sqrt{1-x^{2}} \hspace{0.2cm} (x\leq
1),
\end{array}
\right.
\end{equation}
where i=1,2, $E_{1}=E_{h}|\sin (\alpha _{n}-\alpha )|$ and
$E_{2}=E_{h}|\cos (\alpha _{n}-\alpha )|$. The energy $E_{1}$ is
for nodes close to the field direction, while $E_{2}$ is for nodes far
from it. $N_{0}$ is the normal-state density of states.

When the magnetic field is along a nodal direction, $\alpha
=\alpha _{n}+m\pi /2$ where $m$ is an integer; the two nodal
positions perpendicular to the field contribute fully, while the
two parallel to the field do not contribute at all because the
Doppler shift is an inner product between the superfluid velocity
$\mathbf{v}_{s}$ and the quasiparticle momentum $\hbar \mathbf{k}_{F}$. When the
field is along a maximal gap direction $(\alpha =\alpha
_{n}+(2m+1)\pi /4)$, in contrast, all four nodal points contribute
equally to the DOS with a factor of $\sqrt{2}$ less than the full
contribution. When we add all the contributions, the total DOS has
minima along nodal directions and maxima along maximal gap
directions. At $T = 0$~ K, the oscillation is cusped with a contrast
of $1/\sqrt{2}$ between the minima and the maxima and is
independent of the magnetic field. At a finite temperature,
however, the sharp contrast is washed out due to thermal effects,
leading to more rounded minima and the DOS oscillation now depends
on the magnetic field.

The oscillation amplitude depends on the gap geometry and
dimensionality of the superconductors. The Vekhter \textit{et al.} model
above assumes the Fermi momentum of nodal quasiparticles (nqp's) to be restricted to the nodal plane. A more realistic calculation, therefore, is to allow nqp Fermi momentum out of the nodal planes, which decreases the amplitude of the angular variation.\cite{won01} We account for the 3D effect by
replacing the Doppler-related energies E$_{1}$ and E$_{2}$ by
$E_{1}^{3D}=E_{h}[\sin ^{2}(\alpha _{n}-\alpha )+\cos ^{2}\theta
]^{1/2}$ and $E_{2}^{3D}=E_{h}[\cos ^{2}(\alpha _{n}-\alpha )+\cos
^{2}\theta ]^{1/2}$ respectively and integrate the DOS over polar
angle $\theta $:
\begin{equation}
N(w,h,\alpha )=\frac{1}{2\pi }\int_{0}^{2\pi }\frac{1}{2}[N_{1}(w,h,\alpha
,\theta )+N_{2}(w,h,\alpha ,\theta )]d\theta .
\end{equation}
The qualitative features are the same as those in the 2D case except that
the oscillation amplitude decreases from 41 \% to 6 \% in 3D
superconductors at $T=0$ K. The 3D effect was indeed confirmed
experimentally in the field-angle heat capacity measurement of
YNi$_{2}$B$_{2}$C.\cite{tuson03} We note that a similar effect is predicted for an (s+g)-wave superconductor.\cite{maki03}

\begin{figure}[tbp]
\centering  \includegraphics[width=8.5cm,clip]{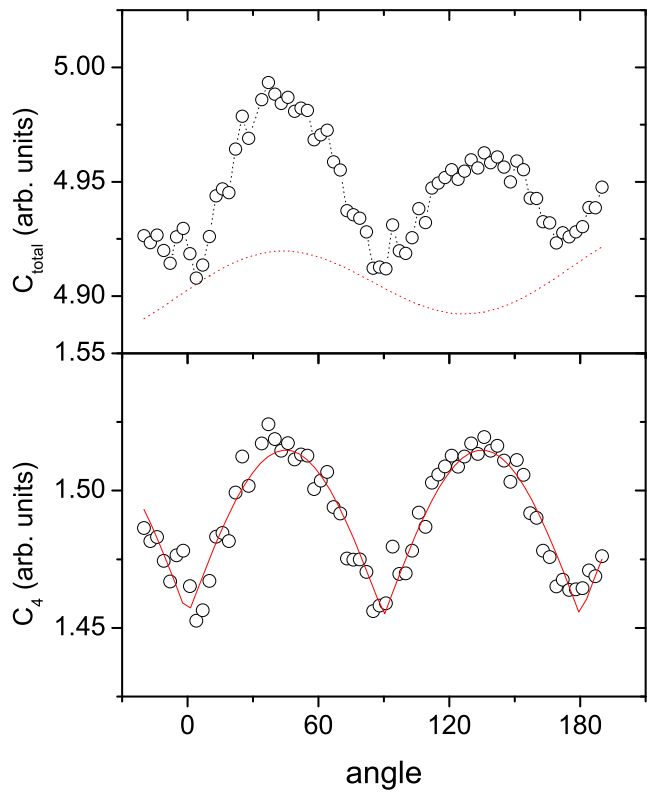}
\caption{Field directional dependence of the heat capacity of sample~A at 2 K and 0.6~T. The field angle $\protect\alpha$ is measured with respect to the a-axis. The top panel shows the total heat capacity (open circles) and the 2-fold
component, $C_{2}(\protect\alpha)$ relative to baseline of 4.9 (dashed
line). The bottom panel shows the same data after subtraction of the 2-fold
and lattice contributions, $C_{4}=C_{total}-C_{2}(\protect\alpha)-C_{0}$. The data were best described by a cusped function, $C_{4}(\alpha, H
)=c_{4}(H)(1+\Gamma |\sin 2\protect\alpha|)$ with $c_{4}=1.45$ and $\Gamma=0.0415$. \label{f6}}
\end{figure}
The upper panel of Fig.~6 shows the field-directional angular dependence of
the total heat capacity at 2~K and 0.6~T, where the transverse magnetic field was rotated within the basal plane of LuNi$_{2}$B$_{2}$C. The total heat capacity consists of constant, 2-fold, and 4-fold contributions: $C_{total}(\alpha, H)=C_{0}+C_{2}(\alpha, H)+C_{4}(\alpha, H)$. The field-independent constant $C_{0}$ is due to nonmagnetic contributions, such
as the lattice heat capacity and thermally excited nqp's, and is determined
experimentally from $C(T)$ at 0~T. The 2-fold contribution $C_{2}(\alpha, H )$ comes
from our experimental setup and has a functional form of $c_{2}\cos 2\alpha $. The Au/Fe and chromel thermocouple wire is a major source of this contribution, but
misalignment of the basal plane of the sample against the field direction
could also lead to the 2-fold component because of the anisotropy between
in-field ab-plane and c-axis heat capacities. The dashed line in the upper
panel of the Fig.~6 is $C_{2}(\alpha, H)$; the 2-fold signal is about 40~$\%$ of
the 4-fold component at 0.6~T and increases with magnetic field. 

The lower panel of the Fig.~6 shows the 0.6~T data after subtraction of
the 2-fold and the lattice components : $C_{4}(\alpha, H)=C_{total}(\alpha, H )-C_{2}(\alpha, H)-C_{0}$. The 4-fold variation is clearly seen. To make a quantitative analysis, we fit our data to $C_{4}(\alpha, H )=c_{4}(H)(1+\Gamma |\sin 2\alpha |)$. The coefficient $c_{4}(H)$ and the angular contrast $\Gamma $ were treated as field dependent fitting parameters. The solid line in the lower panel of Fig.~6 shows the fit with $\Gamma =0.0415$ at 0.6~T. The sharp minima along $<100>$ indicate that there exist gap minima or nodal structures along those directions. The nodal positions and the small oscillation amplitudes are consistent with previous reports of YNi$_{2}$B$_{2}$C.\cite{izawa02,tuson03} The field-angle heat capacities of sample~C and~N at low fields are essentially the same as that of sample~A.
\begin{figure}[tbp]
\centering  \includegraphics[width=8.5cm,clip]{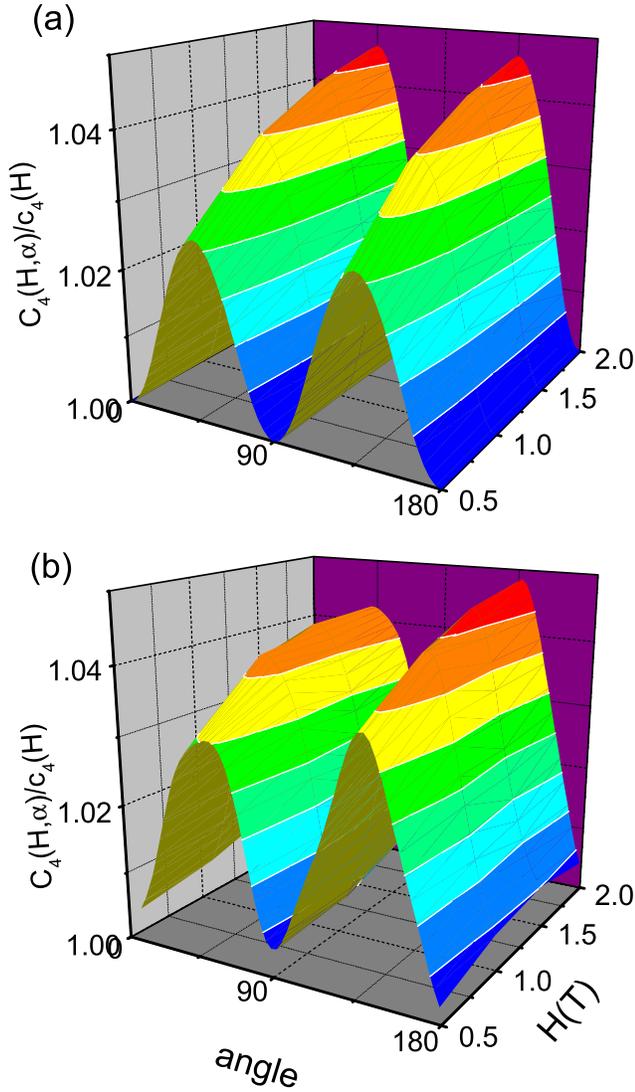}
\caption{The 3D surface plots describe magnetic field-angle
(x-axis) and field-intensity (y-axis) dependence of the in-plane
specific heat ratio at 2.5~K: $C_{4}(H,\alpha)/c_{4}(H)$. Fig.~(a)
is a numerical calculation of the specific heat due to nodal
quasiparticles in 3D system and Fig.~(b) is the experimental data
of the single crystal sample~C. The experimental data were FFT low-pass filtered with a cut-off frequency of 0.03~Hz with angles being considered as time. Both plots (a) and (b) were constructed with 0.5, 1, 1.5, and 2~T data sets. \label{f7}}
\end{figure}

Figure~7 gives 3D surface plots of the field-angle heat capacity of sample~C at 2.5 K. The x-axis is the in-plane field direction $\alpha$ with respect to the crystal axes and the y-axis is the applied magnetic field intensity in tesla. The z-axis is the four-fold heat capacity normalized by the Doppler-induced heat
capacity: $C_{4}(H,\alpha) / c_{4}(H)$. Figure~7(a) is a numerical calculation of the 3D nodal quasiparticle theory \cite{vekhter99,won01} and Fig.~7(b) are the experimental results of sample~C. In the quasiparticle theory, the only adjustable parameter is the Doppler-related energy $E_{h}(v^{*})$ (see Eq.~1).
In the numerical calculation, we used $a=1$, the gap maximum $\Delta =1.76k_{B}T_{c},$ and the average Fermi velocity $v^{*}=\sqrt{v_{a}v_{c}}=1.3\times 10^{7}$ cm/sec. The extracted Fermi velocity is similar to that from the band structure calculation ($\sqrt{v_{a}v_{c}}\sim 1.2\times 10^{7}$ cm/sec).\cite{kogan97} The 3D plots of sample~N show similar behavior to that of sample~C requiring a larger Fermi velocity, i.e., $v^{*}=1.5\times 10^{7}$ cm/sec.\cite{thesis} The larger value may be due to the uncertainty in determining the field-induced heat capacity. 

In field-angle heat capacity measurements, there are three relevant energy scales, which are the Doppler energy scale $E_{h}$, the paramagnetic energy $\mu H$, and the thermal energy of $k_{B}T$. In the mixed state, the paramagnetic energy is of order 1~K and the Doppler energy scale is $\sim 10$~K. The latter is much higher than the experimental temperature 2 K. Therefore, the Doppler effect on the nodal quasiparticles is dominant over the thermal and the paramagnetic effects in this regime, indicating that the 3D nodal quasiparticle theory of the Doppler effect is applicable as was evident from the reasonable agreement between experiments and the numerical calculations.

\begin{figure}[tbp]
\centering  \includegraphics[width=8.5cm,clip]{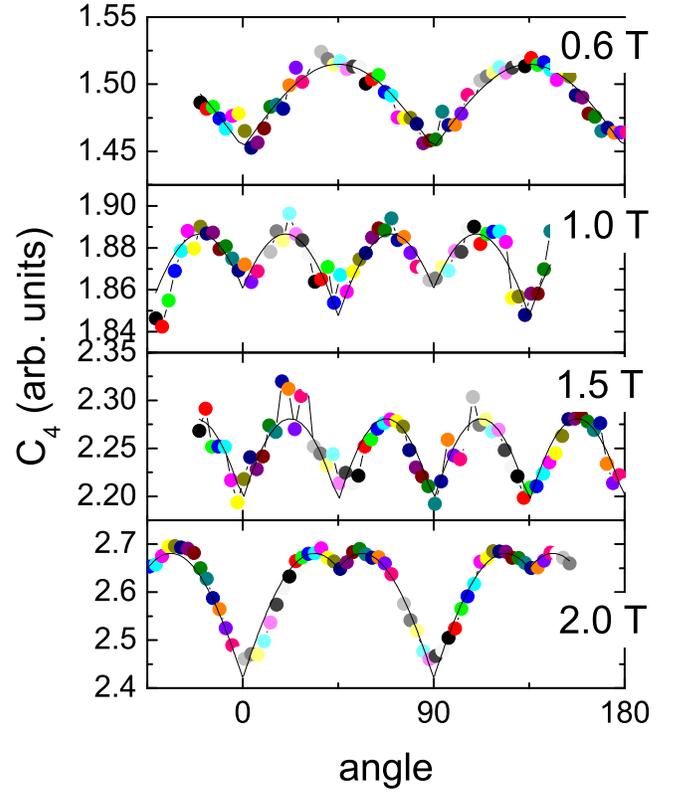}
\caption{Angle-resolved heat capacity of sample~A at 2~K and a $4+4$ pattern (solid line). The magnetic fields are 0.6, 1, 1.5, and 2~T, from top to bottom panel. \label{f8}}
\end{figure}
Figure~8 shows the heat capacity of sample~A as a function of field angle at 2~K and 0.6, 1.0, 1.5, and 2.0~T from top to bottom panel. At 1~T, surprisingly, additional minima develop along $<110>$, producing two sets of fourfold patterns or 8-fold, an effect not observed in either sample~C or~N with higher $T_{c}$ 's. The crossover field from the fourfold to the $(4+4)$ pattern of sample~A lies between 0.6 and 1~T, which is also the point where the heat capacity of sample~A deviates from the $\sqrt{H}$ dependence (see Fig.~5). With
increasing field, the splitting gradually disappears and the
field-angle heat capacity recovers its fourfold pattern above 4~T. We also
measured the field-angle heat capacity of sample~A at 4~K to check if the
anomalous peak splitting persists at higher temperatures (Fig.~9). The fourfold pattern now persists to 1~T, evolving into two sets of
fourfold patterns above 2~T. The 4~T data at 4~K have a shape similar to the
2~T data at 2~K. The crossover field $H_{s1}$ increases with increasing
temperature.
\begin{figure}[tbp]
\centering  \includegraphics[width=8.5cm,clip]{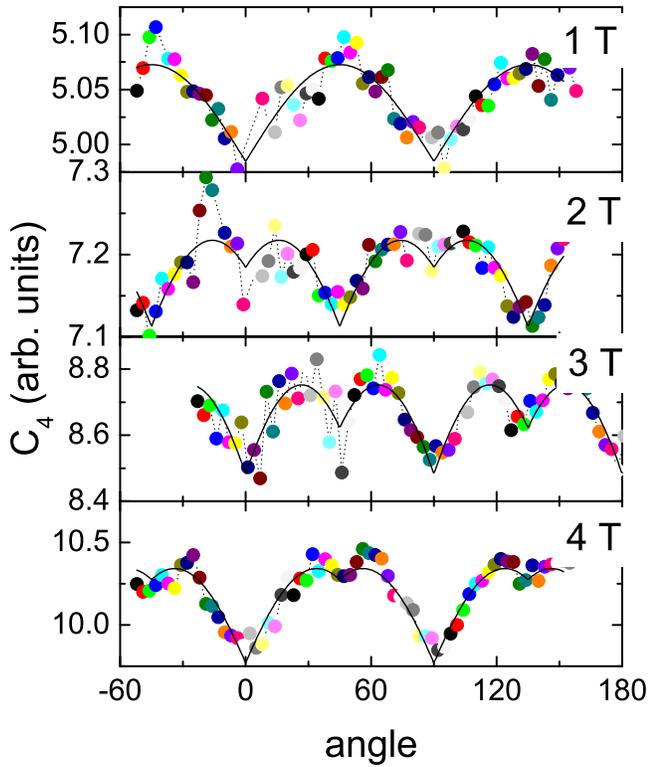}
\caption{Angle-resolved heat capacity of sample~A at 4~K and a $4+4$ pattern (solid line). The applied magnetic fields are 1, 2, 3, and 4~T, from top to bottom panel. \label{f9}}
\end{figure}

It is worth noting that the anomalous 8-fold pattern occurs only at sample~A, which has half the electronic mean free path of the sample~N, while the $T_{c}$ is slightly decreased. According to the nonlocal theory by Kogan \textit{et~al.},\cite{kogan97} the hexagonal-to-square FLL transition depends on the electronic mean free path $l$ and the superconducting coherence length $\xi$. Gammel \textit{et~al.} found that a mere 9\% of Co doping onto the Ni site in Lu1221 can make the FLL transition field 20 times higher than that of the pure matrix for $H \parallel [001]$.\cite{gammel99} The FLL transition field for pure Lu1221 is around 0.2~T,\cite{eskildsen01} which is well below our measurement range, suggesting the nonlocal effects would not influence the field-angle heat capacity of purer samples. In contrast, the disorder in sample~A increases the transition field to at least twice higher than that of sample~N, \cite{gammel99} i.e., to a field relevant in the field-angle heat capacity measurement of sample~A.

When the magnetic field is rotated within the \textit{ab}-plane, the transition field may differ with different field directions because of the different nonlocal
range, i.e., $\xi /l$. The two different transition fields can be
characterized by $H_{s1}$ for $<110>$ and $H_{s2}$ for $<100>$. As a magnetic field rotates within the \textit{ab}-plane for $H_{s1}\leq H\leq H_{s2}$
, the FLL will experience a structural change (or distortion), i.e.
hexagonal for $H\parallel \lbrack 100]$ and square for $H\parallel \lbrack
110]$. Since the borocarbides have nodes on the Fermi surface, the DOS will
differ depending on the FLL structure.\cite{ichioka99} The negative deviation from the $\sqrt{H}$ above $H_{s1}$ in the heat capacity of sample~A (see Fig.~5b) also indicates that the DOS of the hexagonal FLL is larger than that of the square FLL. In addition to the anisotropic gap modulation,\cite{tuson03} the DOS will oscillate due to the FLL anisotropy, leading to a (4+4)-fold pattern.

\begin{figure}[tbp]
\centering  \includegraphics[width=8.5cm,clip]{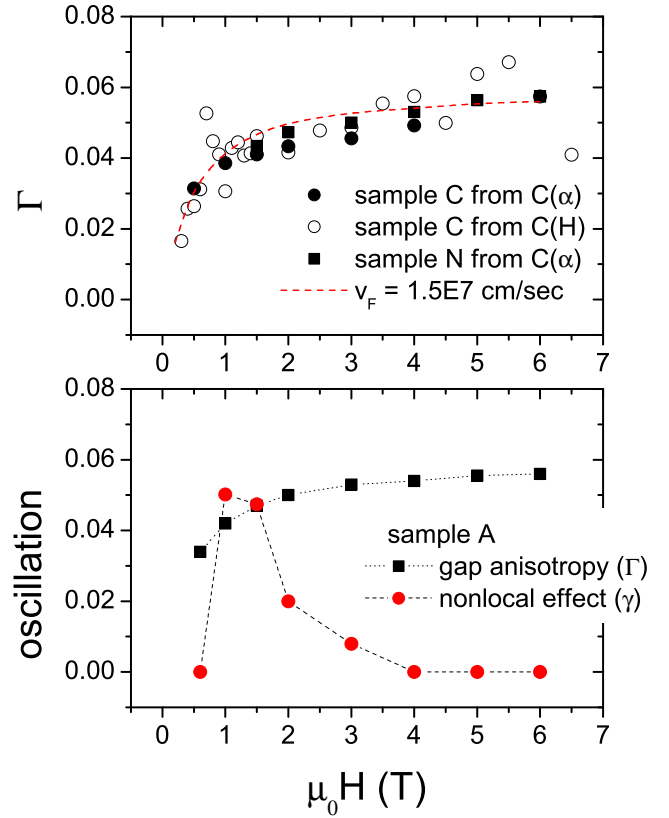}
\caption{Oscillation amplitude $\Gamma$ of sample~C (circles) and sample~N (squares). The dashed line is from the 3D nodal quasiparticle theory with $v_{F}=1.5\times 10^{7}$ cm/sec. The bottom panel contrasts the oscillation amplitude due to the nonlocal effects (circles) and gap anisotropy (squares).$^{38}$ \label{f10}}
\end{figure}
When $H<H_{s1}$ or $H>H_{s2}$, the field-induced parameter $c_{4}$ can be considered as constant and the DOS oscillation is just from the gap anisotropy. When $H_{s1}<H<H_{s2}$, in contrast, $c_{4}$ is not independent of field direction any more, but oscillates due to the anisotropic FLL transition field: $c_{4}=p1(1+\gamma |sin2(\alpha -45)|)$. $c_{4}$ has maxima along $<100>$ and minima along $<110>$. Then, the field-angle heat capacity can be written as 
\begin{equation}
C_{4}(\alpha )=p1(1+\gamma |sin2(\alpha -45)|)(1+\Gamma |sin2\alpha |),
\end{equation}
where $p1$ is a field-dependent fitting parameter. The value $\Gamma $ represents
the oscillation due to gap anisotropy in pure samples (see Fig.~10). The
nonlocal effects give rise to a 45$^{\circ }$-shifted 4-fold pattern and are
accounted for by $\gamma $. The solid lines in Fig.~8 and~9 are least square
fits of Eq.~(4) and represent the data very well. The oscillations due to the nonlocal effects ($\gamma $) and the gap anisotropy ($\Gamma $) at 2~K are compared as a function of magnetic field at the bottom panel of
Fig.~10. The FLL effect $\gamma $ increases sharply above 0.6~T and decreases
gradually to zero at 4~T, indicating that the low field corresponds to $H_{s1}$ and the high field to $H_{s2}$. It is interesting to note that Miranovic \textit{et al.} came to a similar conclusion by solving quasi-classical Eilenberger equation for nodal superconductors with anisotropic Fermi velocity.\cite{miranovic04}

The $H-T$ phase diagram of the disordered sample~A is shown in Fig.~11. $H_{s1}$ is the FLL transition field for $H \parallel [110]$ and $H_{s2}$, for $H \parallel [100]$. The increase in the transition fields with increasing temperature is consistent with the nonlocal prediction where the nonlocal range decreases with raising $T$.\cite{kogan97} The difference in the FLL transition fields for $H \parallel [100]$ and $H \parallel [110]$ can be due to the upper critical field $H_{c2}$ anisotropy between the two directions. Since the effective nonlocal range is proportional to the coherence length $\xi$, the $H_{c2}$ anisotropy leads to a higher FLL transition field for $H \parallel [100]$ than that for $H \parallel [110]$. The FLL transition occurs when the intervortex distance is comparable to the effective nonlocal radius. The observed ratio $H_{s2}/H_{s1}\approx 4$ is larger than the predicted ratio 2,\cite{kogan97} but is smaller than the reported ratio of 10 in YNi$_{2}$B$_{2}$C.\cite{sakata00} The difference between experiments and theory may attest to the need that we take into account both the nonlocality and the anisotropic gap nature of the borocarbides.
\begin{figure}[tbp]
\centering  \includegraphics[width=8.5cm,clip]{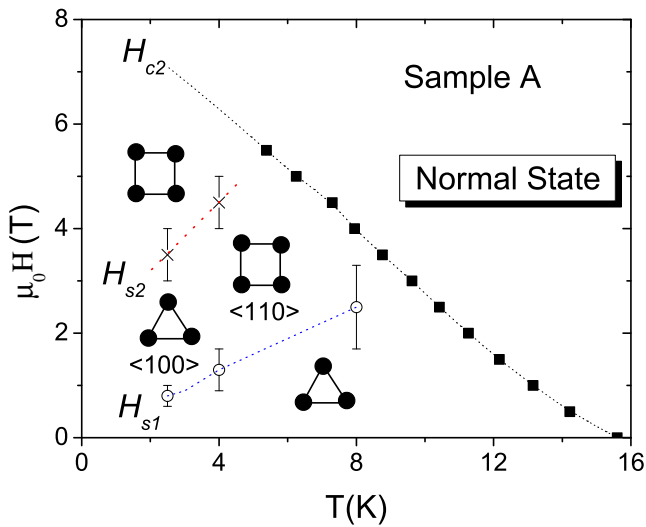}
\caption{$H-T$ phase diagram of sample~A with lowest $T_{c}$. $H_{s1}$ is the FLL transition for $H \parallel$[110] and $H_{s2}$ for $H \parallel$[100]. The dotted lines are guides to the eyes. Triangles and squares were sketched to show the corresponding FLL.$^{38}$}
\end{figure}

\section{Concluding Remarks}

We reviewed the first direct evidence for a variation of the DOS of nodal quasiparticles (nqp's) of unconventional superconductors Lu(Y)Ni$_{2}$B$_{2}$C. The four-fold field-angle oscillation in the heat capacity of the borocarbides is due to the nqp's experiencing a field-induced Doppler energy shift. The 3D superconductivity of the borocarbides is reflected in the small oscillation amplitude of 4~\%, not 40~\% expected for 2D superconductors. A dramatic change has also been observed in a slightly disordered Lu1221. A fourfold pattern in $C(\alpha)$ changed to an eightfold or two sets of fourfold pattern at the FLL transition field, evidencing that both nonlocal effects and an anisotropic superconducting gap coexist in the borocarbides.

Even though it is clear that there are nodes along $<100>$ directions in the borocarbides, the type of nodes has yet to be answered. Based on a theoretical calculation,\cite{maki02} it is asserted that the rapid decrease of the oscillation amplitude of the angle-dependent thermal conductivity with polar angle $\theta$ is due to a point node gap located along $[100]$.\cite{izawa02} Miranovic \textit{et al.}, however, claimed that the oscillation amplitude decreases sharply with $\theta$ for both vertical line node and the $s+g$ point node cases, suggesting that the conical field rotation experiments may not be able to distinguish the two cases.\cite{miranovic04} In order to determine the precise gap structure, instead, they suggested the field dependence of the oscillation amplitudes at low fields. The oscillation monotonically decreases with increasing field for point nodes, while it has a maximum for line nodes. We plan to measure conical field-angle heat capacity and extend our measurements to lower fields, which could shed more light on the controversy of the nature of nodes. A systematic study on disorder effects through Co doping on Ni site, LuNi$_{2-x}$Co$_{x}$B$_{2}$C, will clearly elucidate the correlation between nonlocal effects and anisotropic gap effects. 

It is gratifying that other field-angle heat capacity have been followed after our pathbreaking work.\cite{tuson03} Deguchi \textit{et al.} applied the technique to the spin-triplet superconductor Sr$_{2}$RuO$_{4}$ and found a fourfold oscillation for $H \perp c-$axis, providing decisive information on the multiband superconductivity and gap structures.\cite{deguchi04} Aoki \textit{et al.} have studied the heavy-fermion superconductor CeCoIn$_{5}$ and found a clear fourfold oscillation in $C(\alpha)$ with minima along $<100>$, suggesting the presence of line nodes at those directions.\cite{aoki04} Their interpretation of the superconducting gap symmetry as $d_{xy}-$wave, however, contradicts that ($d_{x^{2}-y^{2}}$) by thermal conductivity data.\cite{izawa01-2} It is interesting to note that the angle-resolved specific heat directly measures the zero-energy density of states (ZEDOS). On the other hand, the thermal conductivity data $\kappa (\alpha)$ necessarily involve both the ZEDOS and the quasiparticle scattering time $\tau (\alpha)$,\cite{salamon95} making it difficult to identify the gap node direction in some cases.

Finally, we mention that the angle-resolved specific heat measurement opened a new venue to better understanding unconventional superconductivity. Even though only a few classes of unconventional superconductors have been studied by this technique,\cite{tuson03,tuson04,aoki04,deguchi04} new features on their superconducting properties have been revealed, indicating a whole new aspect of superconductivity may await to be found.

\section*{Acknowledgments}
Work at Los Alamos was performed under the auspices of the U.S. Department of Energy. Work at University of Illinois was supported by NSF Grant No. DMR 99-72087. X-ray measurements were carried out in the Center for Microanalysis of Materials, University of Illinois, which is partially supported by the U.S Department of Energy under grant DEFG02-91-ER45439. TP acknowledges benefits from discussion with Joe D. Thompson.


\begin{thebibliography}{0}
\expandafter\ifx\csname natexlab\endcsname\relax\def\natexlab#1{#1}\fi
\expandafter\ifx\csname bibnamefont\endcsname\relax
  \def\bibnamefont#1{#1}\fi
\expandafter\ifx\csname bibfnamefont\endcsname\relax
  \def\bibfnamefont#1{#1}\fi
\expandafter\ifx\csname citenamefont\endcsname\relax
  \def\citenamefont#1{#1}\fi
\expandafter\ifx\csname url\endcsname\relax
  \def\url#1{\texttt{#1}}\fi
\expandafter\ifx\csname urlprefix\endcsname\relax\def\urlprefix{URL }\fi
\providecommand{\bibinfo}[2]{#2}
\providecommand{\eprint}[2][]{\url{#2}}

\end{thebibliography}


\begin{thebibliography}{0}
\bibitem{annett96} J. Annett, N. Goldenfeld, and A. J. Leggett, {\it Physical Properties of High Temperature Superconductors vol. 5}, ed. D. M. Ginsberg (World Scientific, Singapore, 1996).

\bibitem{sigrist91} M. Sigrist and K. Ueda, {\it Rev. Mod. Phys} {\bf 63} (1991) 239.

\bibitem{yip92} S. K. Yip and J. A. Sauls, {\it Phys. Rev. Lett.} {\bf 69} (1992) 2264.

\bibitem{maeda95} A. Maeda, Y. Lino, N. Motohira, K.Kishio, and T. Fukase, {\it Phys. Rev. Lett.} {\bf 74} (1995) 1202.

\bibitem{maeda96} A. Maeda {\it et al.}, {\it J. Phys. Soc. Jpn.} {\bf 65} (1996) 3638.

\bibitem{bhattacharya99} A. Bhattacharya {\it et al.}, {\it Phys. Rev. Lett.} {\bf 82} (1999) 3132.

\bibitem{carrington99} A. Carrington, R. W. Giannetta, J. T. Kim, and J. Giapintzakis, {\it Phys. Rev.} {\bf B59} (1999) R14173.

\bibitem{bidinosti99} C. P. Bidinosti, W. N. Hardy, D. A. Bonn, and R. Liang, {\it Phys. Rev. Lett.} {\bf 83} (1999) 3277.

\bibitem{li98} M. R. Li, P. J. Hirschfeld, and P. Wolfle, {\it Phys. Rev. Lett.} {\bf 81} (1998) 5640.

\bibitem{hirschfeld98} C. Kubert and P. J. Hirschfeld, {\it Phys. Rev. Lett.} {\bf 80} (1998) 4693.

\bibitem{salamon95} M. B. Salamon, F. Yu, and V. N. Kopylov, {\it J. Supercon.} {\bf 8} (1995) 449.

\bibitem{yu95} F. Yu, M. B. Salamon, and A. J. Leggett, {\it Phys. Rev. Lett.} {\bf 74} (1995) 5136.

\bibitem{aubin97} H. Aubin, K. Behnia, and M. Ribault, {\it Phys. Rev. Lett.} {\bf 78} (1997) 2624.

\bibitem{izawa01} K. Izawa {\it et al.}, {\it Phys. Rev. Lett.} {\bf 86} (2001) 1327.

\bibitem{izawa02} K. Izawa {\it et al.}, {\it Phys. Rev. Lett.} {\bf 89} (2002) 137006.

\bibitem{vekhter99} I. Vekhter, P. J. Hirschfeld, J. P. Carbotte, and E. J. Nicol, {\it Phys. Rev.} {\bf B59} (1999) R9023.

\bibitem{moler94} K. A. Moler {\it et al.}, {\it Phys. Rev. Lett.} {\bf 73} (1994) 2744.

\bibitem{wang01} Y. Wang, B. Revaz, A. Erb, and A. Junod, {\it Phys. Rev.} {\bf B63} (2001) 094508.

\bibitem{whelan00} N. D. Whelan and J. P. Carbotte, {\it Phys. Rev.} {\bf B62} (2000) 14511.

\bibitem{won01} H. Won and K. Maki, {\it Europhys. Lett.} {\bf 56} (2001) 729.

\bibitem{kraftmakher02} Y. Kraftmakher, {\it Phys. Repts.} {\bf 356} (2002) 1.

\bibitem{sullivan68} P. Sullivan and G. Seidel, {\it Phys. Rev.} {\bf 173} (1968) 679.

\bibitem{thesis} T. Park, Ph.D. thesis, University of Illinois, Urbana-Champaign (2003).

\bibitem{cava94} R. J. Cava {\it et al.}, {\it Nature (London)} {\bf 367} (1994) 252.

\bibitem{mazumdar93} C. Mazumdar {\it et al.}, {\it Solid State Commun.} {\bf 87} (1993) 413.

\bibitem{canfield98} P. C. Canfield, P. L. Gammel, and D. J. Bishop, {\it Phys. Today} {\bf 51} (1998) 40.

\bibitem{nohara97} M. Nohara, M. Isshiki, H. Takagi, and R. J. Cava, {\it J. Phys. Soc. Jpn.} {\bf 66} (1997) 1888.

\bibitem{volovik93} G. E. Volovik, {\it JETP Lett.} {\bf 58} (1993) 469.

\bibitem{boaknin01} E. Boaknin {\it et al.}, {\it Phys. Rev. Lett.} {\bf 87} (2001) 237001.

\bibitem{tuson03} T. Park, M. B. Salamon, E. M. Choi, H. J. Kim, and S.-I. Lee, {\it Phys. Rev. Lett.} {\bf 90} (2003) 177001.

\bibitem{yaron96} U. Yaron {\it et al.}, {\it Nature (London)} {\bf 382} (1996) 236.

\bibitem{eskildsen98} M. R. Eskildsen {\it et al.}, {\it Nature (London)} {\bf 393} (1998) 242.

\bibitem{sakata00} H. Sakata {\it et al.}, {\it Phys. Rev. Lett.} {\bf 84} (2000) 1583.

\bibitem{eskildsen2001} M. R. Eskildsen {\it et al.}, {\it Phys. Rev. Lett.} {\bf 86} (2001) 5148.

\bibitem{kogan97} V. G. Kogan {\it et al.}, {\it Phys. Rev.} {\bf B55} (1997) R8693.

\bibitem{gilardi02} R. Gilardi {\it et al.}, {\it Phys. Rev. Lett.} {\bf 88} (2002) 217003.

\bibitem{cheon98} K. O. Cheon {\it et al.}, {\it Phys. Rev.} {\bf B58} (1998) 6463.

\bibitem{tuson04} T. Park {\it et al.}, {\it Phys. Rev. Lett.} {\bf 92} (2004) 237002.

\bibitem{hedo98} M. Hedo {\it et al.}, {\it J. Phys. Soc. Jpn.} {\bf 67} (1998) 272.

\bibitem{ichioka99} M. Ichioka, A. Hasegawa, and K. Machida, {\it Phys. Rev.} {\bf B59} (1999) 184.

\bibitem{miranovic03} P. Miranovic, N.Nakai, M. Ichioka, and K. Machida, {\it Phys. Rev.} {\bf 68} (2003) 052501.

\bibitem{maki03} K. Maki {\it et al.}, {\it Europhys. Lett.} {\bf 64} (2003) 496.

\bibitem{gammel99} P. L. Gammel {\it et al.}, {\it Phys. Rev. Lett.} {\bf 82} (1999) 4082.

\bibitem{eskildsen01} M. R. Eskildsen {\it et al.}, {\it Phys. Rev. Lett.} {\bf 86} (2001) 320.

\bibitem{miranovic04} P. Miranovic, M. Ichioka, K. Machida, and N. Nakai, cond-mat/0409371, unpublished

\bibitem{maki02} K. Maki, P. Thalmeier, and H. Won, {\it Phys. Rev.} {\bf B65} (2002) 140502.

\bibitem{deguchi04} K. Deguchi, Z. Mao, H. Yaguchi, and Y. Maeno, {\it Phys. Rev. Lett.} {\bf 92} (2004) 047002.

\bibitem{aoki04} H. Aoki {\it et al.}, {\it J. Phys.: Condens. Matter} {\bf 16} (2004) L13.

\bibitem{izawa01-2} K. Izawa {\it et al.}, {\it Phys. Rev. Lett.} {\bf 87} (2001) 157002.

\end{thebibliography}
\end{document}